\documentclass[twocolumn,preprintnumbers,amsmath,amssymb]{revtex4}

\usepackage{epsfig}
\usepackage{subfigure}
\usepackage{graphics}
\usepackage{amsmath}

\begin{document}

\title{$D$-mesons propagation in hadronic matter and consequences
on heavy-flavor observables in ultrarelativistic heavy-ion
collisions}

\author{V.~Ozvenchuk}

\email{Vitalii.Ozvenchuk@subatech.in2p3.fr}

\affiliation{%
  Subatech, UMR 6457, IN2P3/CNRS, %
  Universit$\acute{e}$ de Nantes, $\acute{E}$cole des Mines de Nantes, %
  4 rue Alfred Kastler, 44307 Nantes cedex 3, %
  France %
}

\author{J.~M.~Torres-Rincon}

\affiliation{%
  Subatech, UMR 6457, IN2P3/CNRS, %
  Universit$\acute{e}$ de Nantes, $\acute{E}$cole des Mines de Nantes, %
  4 rue Alfred Kastler, 44307 Nantes cedex 3, %
  France %
}

\author{P.~B.~Gossiaux}

\affiliation{%
  Subatech, UMR 6457, IN2P3/CNRS, %
  Universit$\acute{e}$ de Nantes, $\acute{E}$cole des Mines de Nantes, %
  4 rue Alfred Kastler, 44307 Nantes cedex 3, %
  France %
}

\author{L.~Tolos}

\affiliation{%
  Institut de Ci\`encies de l'Espai (IEEC/CSIC), %
  Campus Universitat Aut\`onoma de Barcelona, Facultat de Ci\`encies, %
  Torre C5, E-08193 Bellaterra, %
  Spain %
}

\affiliation{%
  Frankfurt Institute for Advanced Studies, %
  Johann Wolfgang Goethe--Universit\"at, %
  Ruth-Moufang-Str. 1, 60438 Frankfurt am Main, %
  Germany %
}

\author{J.~Aichelin}

\affiliation{%
  Subatech, UMR 6457, IN2P3/CNRS, %
  Universit$\acute{e}$ de Nantes, $\acute{E}$cole des Mines de Nantes, %
  4 rue Alfred Kastler, 44307 Nantes cedex 3, %
  France %
}

\date{\today}

\begin{abstract}
We employ recently published cross sections for $D$-mesons with
hadrons and calculate the drag and diffusion coefficients of
$D$-mesons in hadronic matter as a function of the momentum of
$D$-mesons as well as of the temperature of the medium. Calculating
in our approach the spatial diffusion coefficient, $D_x$, at zero
chemical potential we see a very smooth transition between our
calculations for the hadron gas and the lattice QCD calculations.
Applying the results for the transport coefficients of $D$-mesons in
a Fokker-Planck equation, which describes the evolution of
$D$-mesons during the expansion of a hadron gas created in
ultrarelativistic heavy-ion collisions, we find that the value of
$R_{AA}$ is little influenced by hadronic rescattering, whereas in
the elliptic flow the effects are stronger. We extend our
calculations to the finite chemical potentials and calculate the
spatial diffusion coefficients of $D$-mesons propagating through the
hadronic medium following isentropic trajectories, appropriate at
future FAIR and NICA heavy-ion experiments. For the isentropic
trajectory with $s/\rho_B^{\rm net}=20$ we find a perfect matching
of results for $D$-mesons in hadronic matter and for charm quarks in
partonic matter treated within the DQPM approach.
 \end{abstract}

\maketitle

\section{Introduction}
\label{introduction} A new state of matter, a deconfined phase of
quarks and gluons known as quark-gluon plasma (QGP), has been
created in ultrarelativistic heavy-ion collisions at the
Relativistic Heavy-Ion Collider (RHIC) at Brookhaven National
Laboratory (BNL)~\cite{STAR,PHENIX,BRAHMS,PHOBOS} and at the Large
Hadron Collider (LHC) at the European Organization for Nuclear
Research (CERN)~\cite{ALICE}.

Heavy-flavor (HF) particles have been suggested to study the
properties of the QGP. Heavy quarks (charm and bottom) are produced
in the initial hard nucleon-nucleon scatterings and their initial
momentum distribution can be directly inferred from proton-proton
collisions. Initially the heavy quarks have no elliptic flow. Their
thermal equilibration time  is larger than the QGP lifetime in
heavy-ion collisions (HICs). Thus the interaction between heavy
quarks and the partons of the QGP modify the heavy quark spectra but
do not bring heavy quarks to a thermal equilibrium.

A direct measure of the interaction of the heavy quarks with the QGP
medium created in HICs is the nuclear modification factor, $R_{AA}$,
which is the ratio of the $p_T$ distribution measured in HICs and
the reference $p_T$ distribution from proton-proton collisions
scaled by the number of binary collisions. It is therefore one if a
HIC is just an ensemble of independent $pp$ collisions. The $R_{AA}$
of intermediate- and high-$p_T$ HF mesons and single nonphotonic
electrons originating from the decays of HF mesons, experimentally
measured at
RHIC~\cite{PHENIX_electrons_R_AA_v2,STAR_electrons_R_AA,STAR_Dmeson_R_AA}
and LHC~\cite{ALICE_Dmeson_R_AA,ALICE_electrons_R_AA_v2}, is
significantly below unity. It indicates a substantial energy loss of
high-$p_T$ heavy quarks in the QGP. The observed finite elliptic
flow, $v_2$, of HF mesons and decay
electrons~\cite{PHENIX_electrons_R_AA_v2,STAR_Dmeson_v2,ALICE_electrons_R_AA_v2,ALICE_Dmeson_v2}
is either due to the interactions with light quarks and gluons in
the plasma leading to the partial thermalization of low- and
intermediate-$p_T$ heavy quarks or due to the coalescence at the end
of the deconfined phase.

The large masses of charm ($c$) and bottom ($b$) motivated different
authors to calculate the time evolution of distribution function of
heavy quarks in a plasma by a Fokker-Planck (FP)
equation~\cite{FP_Svetitsky,FP_Hees_Rapp,FP_Moore_Teaney,FP_Mustafa,
FP_Gossiaux_Guiho_Aichelin,L_Hees_Greco_Rapp,FP_Ko_Liu,L_Akamatsu_Hatsuda_Hirano,
FP_Das_Alam_Mohanty,L_Alberico,L_Cao_Qin_Bass}. In such an approach
the drag and diffusion coefficients depend on the temperature of the
expanding plasma and the momentum of the heavy quark. The
instantaneous temperature of the expanding plasma is an input in
these calculations. It may be obtained by a hydrodynamical
description of the expanding plasma. Later it turned out that the
small scattering angle assumption, inherent in a FP approach, is not
completely justified and therefore approaches have been launched
which replace the FP equation by a Boltzmann
equation~\cite{FP_Gossiaux_Aichelin,FP_Gossiaux_Bierkandt_Aichelin}.

In order to exploit the  HF particles as an efficient probe for the
understanding of  the time evolution of the partonic matter the
effects of the interactions of HF hadrons with hadronic matter has
to be understood and the observables have to be corrected for this
effect. Different approaches have been used to study the interaction
of HF mesons in hadronic
matter~\cite{diff_hadron_ChPT,diff_hadron_scat_amp,diff_hadron_Born,diff_hadron_unitarized_1,
diff_hadron_unitarized_2,diff_hadron_unitarized_3,diff_hadron_unitarized_4,diff_hadron_He_Fries_Rapp,diff_hadron_Das}.
This interaction has been usually described in form of drag and
diffusion coefficients. The diffusion coefficient of $D$-mesons was
calculated using an effective theory incorporating both chiral and
heavy-quark symmetries~\cite{diff_hadron_ChPT}. In
Ref.~\cite{diff_hadron_scat_amp} the drag and diffusion coefficients
of $D$-mesons were obtained employing empirical elastic scattering
amplitudes of $D$-mesons with thermal hadrons. The interactions of
$D$-mesons with light mesons and baryons were evaluated using Born
amplitudes~\cite{diff_hadron_Born} and within an unitarized approach
based on effective models~\cite{diff_hadron_unitarized_3}.

In the present paper, we calculate the drag and diffusion
coefficients of $D$-mesons propagating in hadronic matter by
extending the model of Ref.~\cite{diff_hadron_unitarized_3}.
Subsequently  we implement the transport coefficients into our
transport
model~\cite{FP_Gossiaux_Aichelin,FP_Gossiaux_Bierkandt_Aichelin,model_1}.
This model was developed to describe the momentum distribution of
heavy quarks or HF mesons produced in HICs, and to evaluate the
nuclear modification factor and elliptic flow of $D$-mesons as well
as of single nonphotonic electrons originating from the decays of HF
mesons in HICs at $\sqrt{s}=200~{\rm GeV}$ at RHIC. Finally, we
extend our calculations and present the $D$-meson propagation in
hadronic medium following isentropic trajectories, appropriate at
future FAIR and NICA heavy-ion experiments.

The paper is organized as follows. In Sec.~\ref{model} we provide
the detailed description of the model, which is used to propagate heavy
quarks and HF mesons in the bulk medium. We discuss in
Sec.~\ref{hadron_gas} the hadronic cocktail used to calculate the
$D$-meson drag and diffusions coefficients. In
Sec.~\ref{transport_coeff} we then present the results for the
transport coefficients of $D$-mesons and compare these results to
previous studies in this topic. The resulting nuclear modification
factor, $R_{AA}$, and elliptic flow, $v_2$, of $D$-mesons with the
presence of $D$-meson rescattering in the hadronic medium are given
in Sec.~\ref{R_AA_and_v2}. In Sec.~\ref{FAIR} we show the transport
coefficients of $D$-meson at FAIR energies and Sec.~\ref{summary} is
dedicated to the summary and outlook.

\section{The model}
\label{model} In this study we use the Monte Carlo propagation of
heavy quarks and HF
mesons~\cite{FP_Gossiaux_Aichelin,FP_Gossiaux_Bierkandt_Aichelin,model_1},
MC@sHQ, within a $2+1$d fluid dynamically expanding ideal plasma
with smooth initial conditions and an equation of state with a
first-order phase transition~\cite{KH_hydro}.

We initialize the heavy quarks at the original nucleon-nucleon
scattering points according to the $p_T$-distribution from
perturbative QCD (pQCD) results in fixed order plus next to leading
logarithm (FONLL)~\cite{FONLL_1,FONLL_2,FONLL_3}. They are
isotropically distributed in the azimuthal direction and therefore
$v_2$ is initially zero. In coordinate space the initial
distribution of heavy quarks is given by the Glauber calculation.

The local temperature and velocity field determine the interactions
of heavy quarks with locally thermalized partons in the plasma.
They allow us to evaluate the number density and
momentum distribution of the plasma particles. Heavy quarks interact
with plasma partons by either elastic or radiative collisions.
The evolution of heavy quarks is described by the Boltzmann equation,
which is solved by the test particle method, applying Monte Carlo techniques.

The elastic cross sections of heavy quarks with the gluons and light
quarks in the plasma are obtained from pQCD matrix elements in Born
approximation~\cite{FP_Svetitsky,elastic_Born} including an
effective running coupling constant $\alpha_{\rm
eff}(Q^2)$~\cite{running_coupling_1,running_coupling_2,running_coupling_3},
determined from electron-positron
annihilation~\cite{ep_annihilation} as well as nonstrange hadronic
decays of $\tau$ leptons~\cite{tau_decay}. For the gluon propagator
we respectively use a hard thermal loop (HTL)
approach~\cite{HTL_1,HTL_2} and a semi-hard
propagator~\cite{FP_Gossiaux_Aichelin} for small and large
transferred momentum.

%Radiative collisions are calculated via matrix elements from scalar
%QCD~\cite{incoherent_sQCD}. Radiative collisions are not independent
%(Landau-Pomeranchuk-Migdal effect). This is taken into account and
%yields an effective reduction of the radiation
%spectrum~\cite{coherent_reduction}.

The hadronization of heavy quarks occurs when the energy density of
the fluid cell is less than a critical value,
$\varepsilon_c=0.45~{\rm GeV/fm^3}$, (corresponding to a critical
temperature of $T_c=165~{\rm MeV}$). There is a mixed phase (from
$\varepsilon_{\max}=1.65~{\rm GeV/fm^3}$ to $\varepsilon_c=0.45~{\rm
GeV/fm^3}$) in our model at $T_c$, where all interaction rates are
rescaled by a factor of $\varepsilon/\varepsilon_{\rm max}$. The
heavy quarks form hadrons via
coalescence~\cite{FP_Gossiaux_Bierkandt_Aichelin}, predominantly for
low-$p_T$ heavy quarks, or fragmentation~\cite{fragmentation},
predominantly for intermediate- and high-$p_T$ quarks.

Originally it has been assumed in this model that after the
hadronization HF mesons do not
interact~\cite{FP_Gossiaux_Aichelin,FP_Gossiaux_Bierkandt_Aichelin}.
This leads to some tension between the $R_{AA}$ and the $v_2$. It is
one of the purposes of this article to study explicitly the
interaction of $D$-mesons with hadrons in the medium consisting of
light and strange mesons as well as of baryons and antibaryons (as
described in Sec.~\ref{hadron_gas}) and to see whether one can
obtain a better agreement for $R_{AA}$ and $v_2$ simultaneously. For
this purpose we evaluate the drag and diffusion coefficients and use
the FP equation~\eqref{FP_equation} to calculate the modification of
the spectrum as well as of the elliptic flow due to hadronic
interactions. The hadronic medium during the expansion is described
by \cite{KH_hydro}.

%The solution of the FP equation,, is then included to the MC@sHQ model to
%evaluate the impact of hadronic medium on the HF experimental
%observables in HICs at RHIC.

\section{The hadron gas}
\label{hadron_gas} In this section we concentrate on the description
of the hadronic medium employed to evaluate the drag and diffusion
coefficients of $D$-mesons propagating through a hadronic medium. In
ultrarelativistic HICs at RHIC energies the chemical freeze-out of
hadron ratios at a temperature of $T_{\rm ch}\simeq 170~{\rm MeV}$
is significantly earlier than thermal freeze-out of the light
hadrons at $T_{\rm th}\simeq 90~{\rm MeV}$. Thus, in a
thermodynamical description of the cooling process from chemical to
thermal freeze-out, the conservation of the observed particle ratios
(given at the chemical freeze-out) can be achieved by the
introduction of effective chemical potentials for all hadron species
that are not subject to strong decays on the scale of the typical
expansion time~\cite{effective_potential_rapp}.

We simulate the hadron medium, which consists of light mesons
($\pi$, $\eta$, $\rho$, $\omega$, $\eta^{\prime}$, $f_0$, $a_0$,
$\phi$, $h_1$, $b_1$, $a_1$, $f_2$, $f_1$), strange mesons ($K$,
$K^*$, $K_1$), nucleons ($p$, $n$), nuclear resonances [$N(1440)$,
$N(1520)$, $N(1535)$, $N(1650)$, $N(1675)$, $N(1680)$, $N(1700)$]
and $\Delta$-resonances [$\Delta(1232)$, $\Delta(1600)$,
$\Delta(1620)$, $\Delta(1700)$] as well as the corresponding
antibaryons, at various values for the temperature [in a range from
thermal ($90~{\rm MeV}$) to chemical ($170~{\rm MeV}$) freeze-out].
The study of thermal and chemical equilibration of a hadron matter
in HICs have been performed in Ref.~\cite{hadron_matter_Elena} using
a transport model. If we implement the effective chemical potentials
into the thermal hadron distribution functions the particle
densities in the system at given temperature can be calculated as
\begin{equation}
n_i=d_i\int\frac{d^3k}{(2\pi)^3}\frac{1}{e^{[E_i-\mu_i(T)]/T}\pm
1},~(i=\pi,K,...,N,\Delta) \label{particle_density}
\end{equation}
and the total energy density is given by
\begin{equation}
\varepsilon=\sum_id_i\int\frac{d^3k}{(2\pi)^3}\frac{E_i}{e^{[E_i-\mu_i(T)]/T}\pm
1},~~E_i=\sqrt{k^2+m_i^2}, \label{energy_density}
\end{equation}
where $T$ denotes the temperature of the system, $\mu_i$ and $d_i$
stand for the effective chemical potential and degeneracy factor of
$i$-particle, respectively. In Eqs.~(\ref{particle_density},
\ref{energy_density}) upper (lower) signs refer to bosons
(fermions), respectively.

\begin{figure}
\centering
\includegraphics[width=0.5\textwidth]{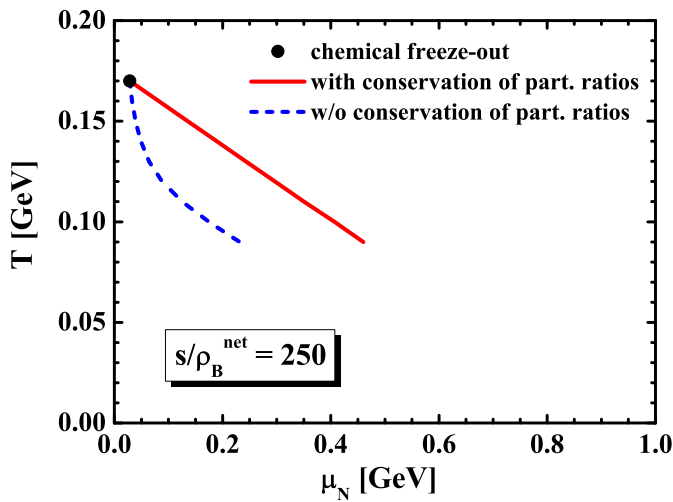}
\caption{Isentropic thermodynamic trajectories at top RHIC energies.
Solid line includes the conservation of particle ratios and dashed
line does not.}\label{isentropic_trajectory}
\end{figure}

\begin{figure*}
\centering \subfigure{
\resizebox{0.48\textwidth}{!}{%
 \includegraphics{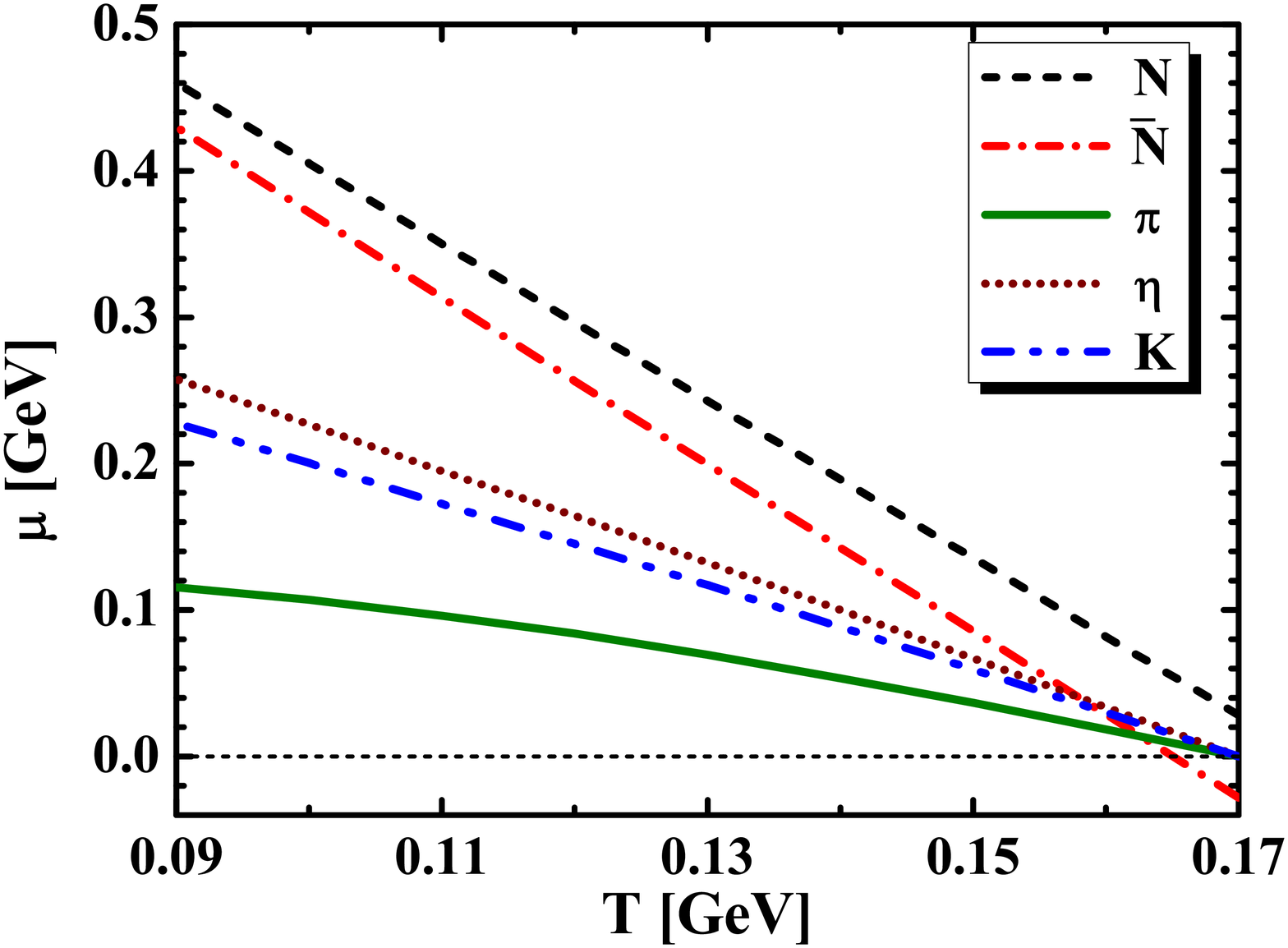}
} } \subfigure{
\resizebox{0.48\textwidth}{!}{%
 \includegraphics{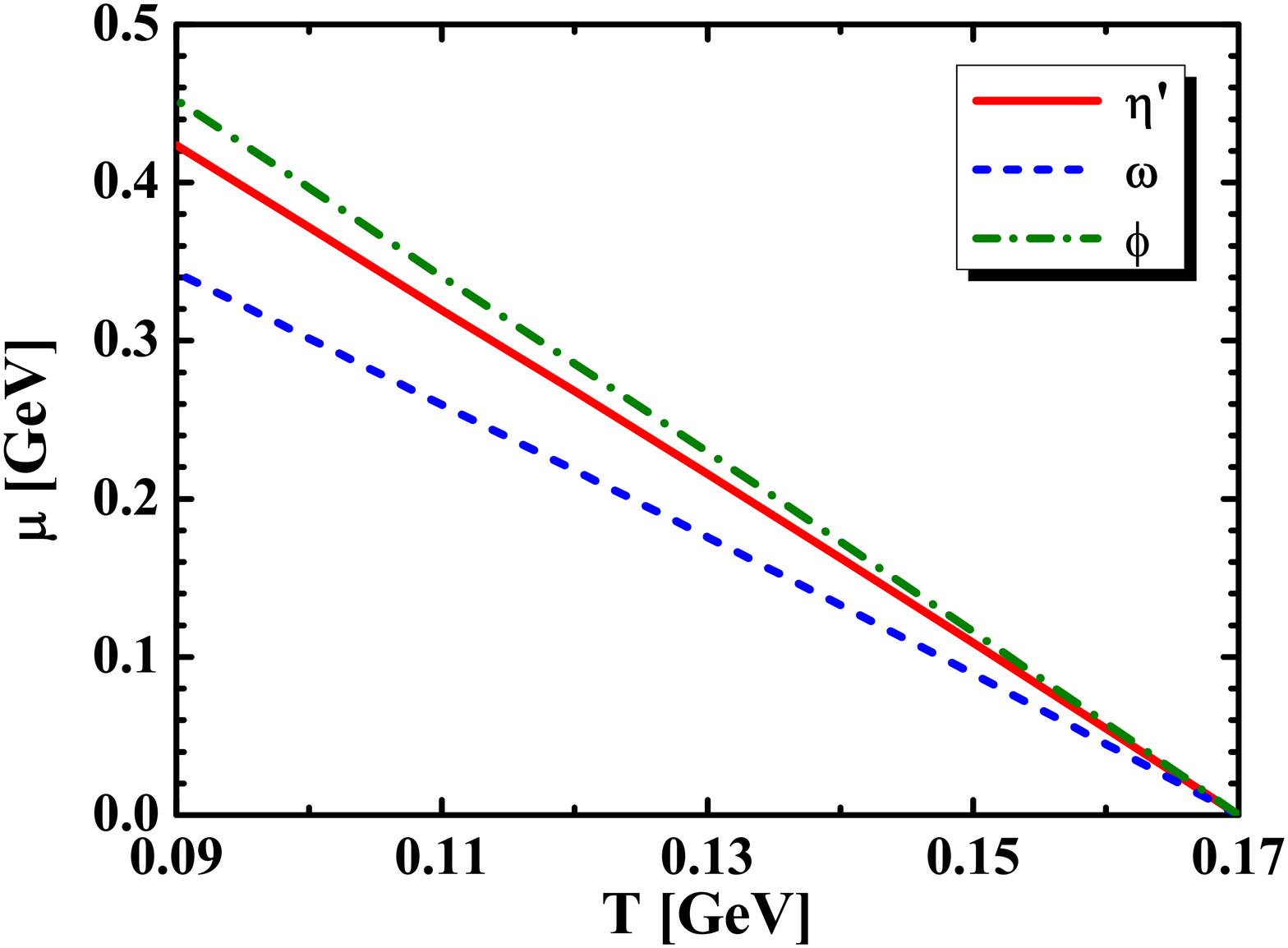}
} } \caption{The temperature dependence of effective chemical
potentials for an isentropic hadronic fireball expansion at top RHIC
energy for different particle species: (left panel) $\pi$ (solid),
$N$ (dashed), $\bar N$ (dash-dotted), $\eta$ (dotted), $K$
(dash-dot-dotted); (right panel) $\eta^{\prime}$ (solid), $\omega$
(dashed), $\phi$ (dash-dotted).} \label{effective_potential}
\end{figure*}

We assume an adiabatic expansion with  a specific entropy (i.e.,
entropy per net baryon) of $S/N_B^{\rm net}=250$ which is a
characteristic value for collisions at the top RHIC energy. This
entropy together with the temperature of chemical freeze-out,
$T_{\rm ch}=170$~MeV, results in a baryon chemical potential of
$\mu_B^{\rm ch}=28.3~{\rm MeV}$. At chemical freeze-out temperature,
$T_{\rm ch}$, all meson effective chemical potentials are equal to
zero, $\mu^{\rm ch}_m=0$. Starting from the chemical freeze-out
point and employing the entropy density
\begin{equation}
s=\mp\sum_id_i\int\frac{d^3k}{(2\pi)^3}[\pm f\ln{f}+(1\mp
f)\ln{(1\mp f)}],
\label{entropy_density}
\end{equation}
as well as the net baryon density
\begin{equation}
\rho_B^{\rm
net}=\sum_{B_i}d_{B_i}\int\frac{d^3k}{(2\pi)^3}[f^{B_i}(\mu_{B_i},T)-f^{\bar
B_i}(\mu_{\bar B_i},T)]
\label{net_baryon_density}
\end{equation}
we can construct a thermodynamic trajectory in the $\mu_N$-$T$
plane for fixed  $s/\rho_B^{\rm net}(T)=250$. The upper (lower)
signs refer to fermions (bosons) and $f$ are the particle thermal
distribution functions. The sum is taken over all particles and over
all baryons in Eqs.~\eqref{entropy_density} and
\eqref{net_baryon_density}, respectively. The resulting trajectory
is shown by the dashed line in Fig.~\ref{isentropic_trajectory}.
Note that, in this case all meson effective chemical potentials are
equal to zero during the evolution, $\mu_m^{\rm eff}(T)=0$, i.e., we
do not conserve the particle ratios during the expanding of the
fireball from chemical to thermal freeze-out.

We come now to the expansion of the hadron gas in which the particle
ratios are conserved. The effective pion number is defined as
\begin{equation}
N_{\pi}^{\rm eff}=V\sum_i N_{\pi}^{(i)}n_i(T,\mu_i),
\end{equation}
where $V$ denotes the volume and $n_i$ stands for the particle
density of $i$-resonance with lifetimes shorter than the typical
fireball lifetime. The effective pion number $N_{\pi}^{(i)}$ of a
given resonance is determined by the number of pions in its decay
modes taking into account the branching ratios, e.g.,
$N_{\pi}^{(\rho)}=N_{\pi}^{(f_0)}=2$,
$N_{\pi}^{(\Delta)}=N_{\pi}^{(K^*)}=1$,
$N_{\pi}^{(N(1440))}=0.65\cdot1+0.35\cdot2=1.35$, etc. The
corresponding effective chemical potentials for those resonances are
defined with the same weighting, e.g.,
$\mu_{\rho}=\mu_{f_0}=2\mu_{\pi}$, $\mu_{\Delta}=\mu_N+\mu_{\pi}$,
$\mu_{K^*}=\mu_K+\mu_{\pi}$, $\mu_{N(1440)}=\mu_N+1.35\mu_{\pi}$,
etc. In analogy, all other effective meson numbers, which have to be
conserved, can be defined for all mesons that are not subject to
strong decays, e.g., for $K$, $\eta$, $\eta^{\prime}$, $\omega$,
$\phi$ mesons. In our calculation we neglect the in-medium
modifications of the spectral function of $\omega$ and $\phi$
mesons. As first pointed out by Rapp in
Ref.~\cite{effective_potential_rapp}, the explicit conservation of
the number of antibaryons significantly affects the composition of
the hadronic expansion at RHIC. This idea was then employed  in the
hadronic equation of state for the hydrodynamical model of Heinz and
Kolb~\cite{KH_effective}, which we use to describe the bulk medium
in the MC@sHQ model. Thus, for the consistency of our calculations,
we also conserve the number of antibaryons by introducing the
effective antibaryon chemical potential, $\mu_{\bar B}^{\rm eff}$,
e.g., $\mu_{\bar N}=-\mu_{N}+\mu_{\bar B}^{\rm eff}$,
$\mu_{\bar\Delta}=-\mu_{\Delta}+\mu_{\bar B}^{\rm eff}+\mu_{\pi}$.
Now starting again from the freeze-out point we can construct an
isentropic trajectory, $s/\rho_B^{\rm net}(T)=250$, keeping also the
ratios of effective stable particle numbers to effective antibaryon
number
\begin{equation}
\frac{N_B^{\rm eff}}{N_{\bar B}^{\rm
eff}},\,\,\,\,\frac{N_{\pi}^{\rm eff}}{N_{\bar B}^{\rm
eff}},\,\,\,\,\frac{N_{\eta}^{\rm eff}}{N_{\bar B}^{\rm
eff}},\,\,\,\,\frac{N_K^{\rm eff}}{N_{\bar B}^{\rm
eff}},\,\,\,\,\frac{N_{\omega}^{\rm eff}}{N_{\bar B}^{\rm
eff}},\,\,\,\,\frac{N_{\eta^{\prime}}^{\rm eff}}{N_{\bar B}^{\rm
eff}},\,\,\,\,\frac{N_{\phi}^{\rm eff}}{N_{\bar B}^{\rm eff}},
\label{particle_ratios}
\end{equation}
fixed in the hadronic evolution toward thermal freeze-out. The
resulting thermodynamic trajectory with the conservation of particle
ratios is presented in Fig.~\ref{isentropic_trajectory} by the solid
line. A fixed specific entropy density together with the
conservation of particle ratios during the fireball evolution lead
to an approximately linear increase of the effective meson chemical
potentials with decreasing temperature, which are shown in
Fig.~\ref{effective_potential} along with nucleon and antinucleon
chemical potentials.

\begin{figure}
\centering
\includegraphics[width=0.5\textwidth]{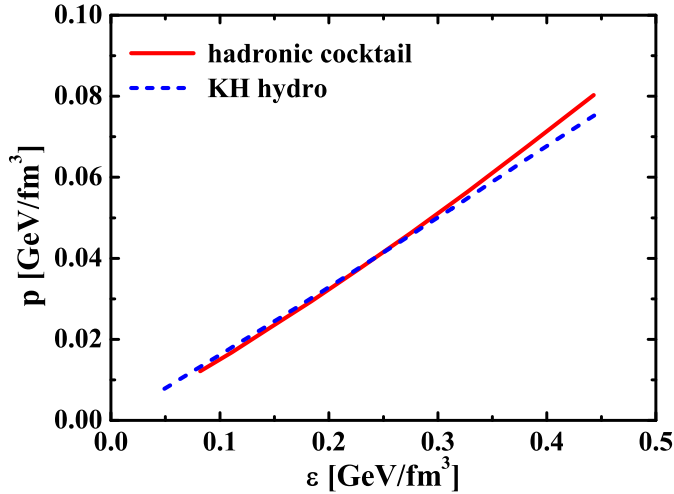}
\caption{The pressure as a function of energy density calculated in
our approach (solid) in comparison to the result obtained from the
hydrodynamical model~\cite{KH_hydro} (dashed).} \label{eos_pressure}
\end{figure}

Before we proceed to the results for $D$-meson transport
coefficients the applicability of the results to the bulk medium in
the MC@sHQ model has to be checked. Thus, we compare the equation of
state used in our calculations to the hadronic equation of state
employed in the hydrodynamical model of Heinz and
Kolb~\cite{KH_hydro}. The pressure in our hadronic cocktail can be
calculated as
\begin{equation}
p=\sum_id_i\int\frac{d^3k}{(2\pi)^3}\frac{k^2}{3E_i}\frac{1}{e^{[E_i-\mu_i(T)]/T}\pm
1},
\label{pressure}
\end{equation}
where the sum is taken over all particles in the system. We present
in Fig.~\ref{eos_pressure} the pressure~\eqref{pressure} as a
function of energy density~\eqref{energy_density} in comparison to
the result from the hydrodynamical model~\cite{KH_hydro}. After the
examination of Fig.~\ref{eos_pressure} we can conclude that the
equations of state are in agreement.

\section{The $D$-meson transport coefficients}
\label{transport_coeff} The propagation of $D$-mesons in the
hadronic medium is evaluated using the FP equation
\begin{equation}
\label{FP_equation}
\frac{\partial f({\bf p},t)}{\partial
t}=\frac{\partial}{\partial p_i}\biggl[A_i({\bf p})f({\bf
p},t)+\frac{\partial}{\partial p_j}B_{ij}({\bf p})f({\bf
p},t)\biggr],
\end{equation}
which describes the time evolution of the distribution of the heavy
quark $f({\bf p},t)$ in an medium characterized by the (time
dependent) drag, $A_i({\bf p})$, and diffusion, $B_{ij}({\bf p})$,
coefficients, which are given by
\begin{equation}
A_i({\bf p})=\int d{\bf k}\,\omega({\bf p,k})\,k_i,
\label{drag_coeff}
\end{equation}
\begin{equation}
B_{ij}({\bf p})=\int d{\bf k}\,\omega({\bf p,k})\,k_ik_j.
\label{diff_coeff}
\end{equation}
$\omega({\bf p,k})$ is the transition rate for a collisions of a
$D$-meson with heat-bath particles with initial ${\bf p}$ and final
${\bf p-k}$ momenta, ${\bf k}$  being the transferred momentum. For
the elastic scattering of a $D$-meson with momentum ${\bf p}$ on a
thermal hadron with momentum ${\bf q}$, the transition rate,
$\omega({\bf p,k})$, can be expressed as
\begin{equation}
\omega({\bf p,k})=g_h\!\int\!\!\!\frac{d{\bf q}}{(2\pi)^3}f_h({\bf
q})v_{\rm rel}\frac{d\sigma}{d\Omega}({\bf p,q\rightarrow k,q+k}).
\end{equation}
Here the  index $h$ stands for the hadrons in the thermal bath, $g_h$
represents the spin-isospin degeneracy factor and $f_h$ is the Bose
or Fermi distribution for thermal hadrons. The relative velocity,
$v_{\rm rel}$, is determined as
\begin{equation}
v_{\rm rel}=\frac{\sqrt{(p\cdot q)^2-(m_Dm_h)^2}}{E_DE_h},
\end{equation}
where $p=(E_D,{\bf p})$ and $q=(E_h,{\bf q})$ are the four-momenta
of the $D$-meson and the thermal hadron, respectively.

\begin{figure}
\centering \subfigure{
\resizebox{0.49\textwidth}{!}{%
 \includegraphics{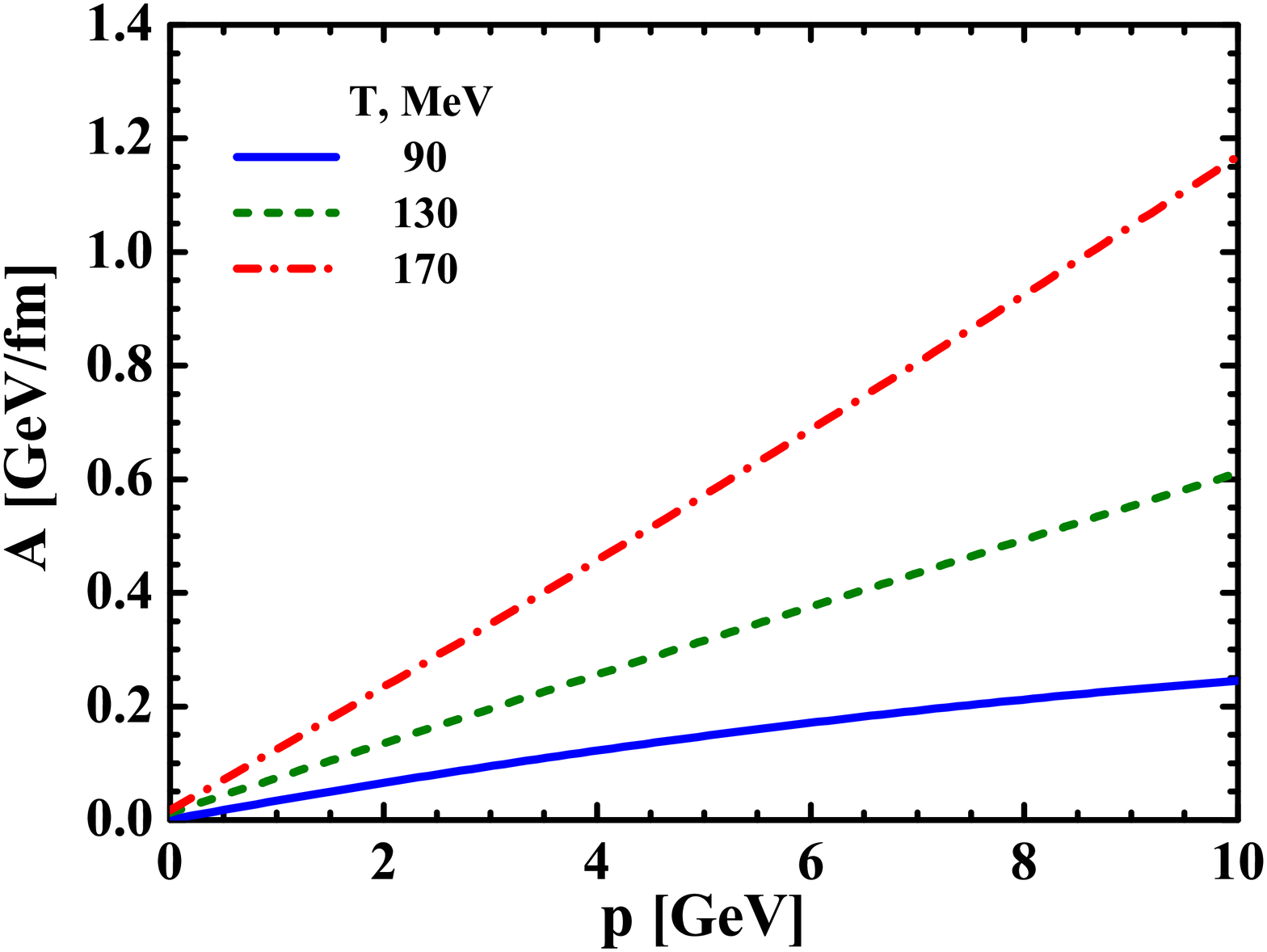}
} } \subfigure{
\resizebox{0.49\textwidth}{!}{%
 \includegraphics{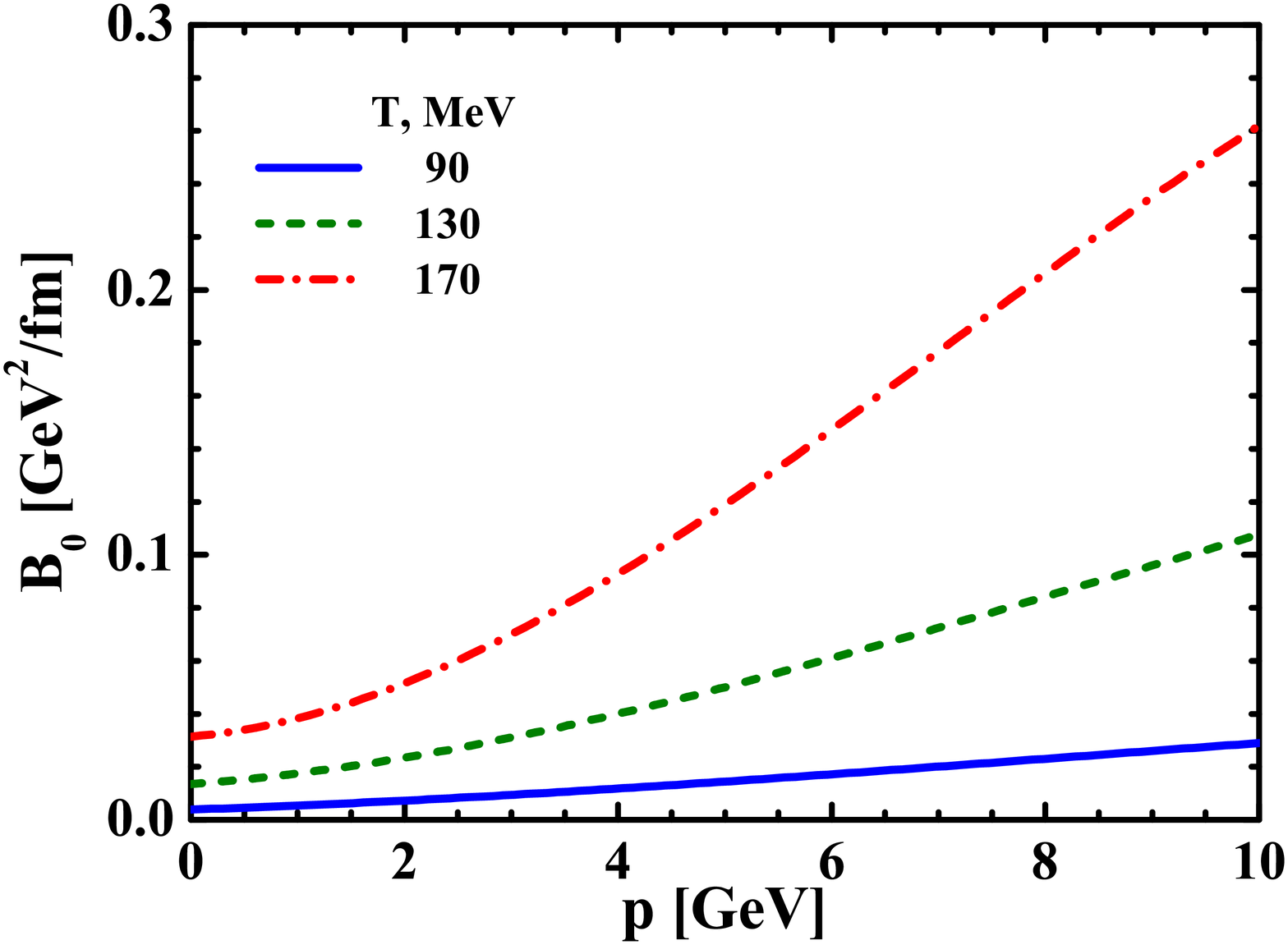}
} } \subfigure{
\resizebox{0.49\textwidth}{!}{%
 \includegraphics{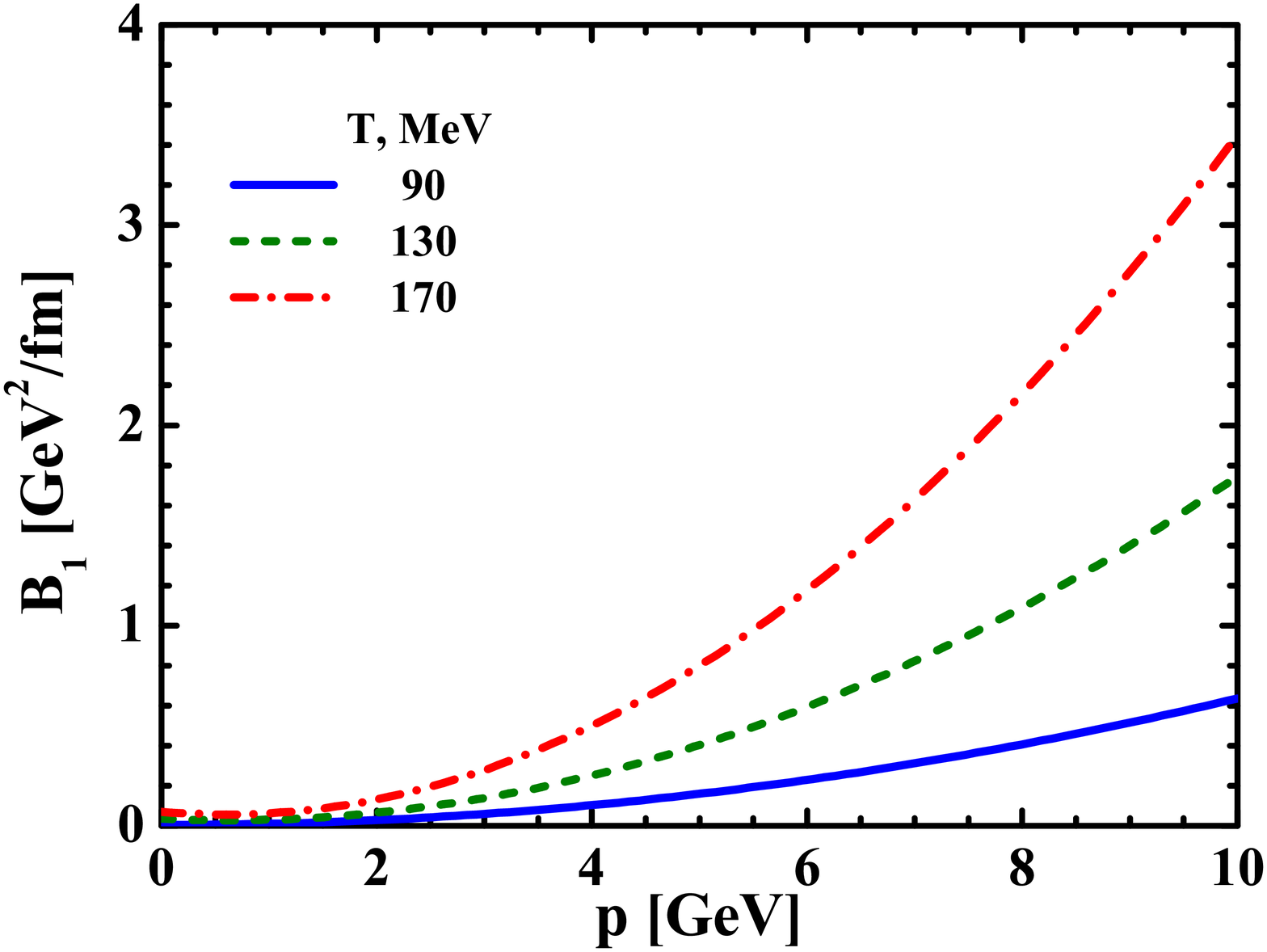}
} } \caption{The transport coefficients as a function of the initial
momentum of the $D$-meson for systems at different temperatures:
drag, $A$ (upper panel); transverse diffusion, $B_0$ (middle panel);
longitudinal diffusion, $B_1$ (lower panel).}
\label{drag_and_diffusion}
\end{figure}

For an isotropic background medium, the drag, $A_i({\bf p})$, and
diffusion, $B_{ij}({\bf p})$, coefficients can be written as
\begin{equation}
A_i({\bf p})=\gamma({\bf p})p_i, \label{drag}
\end{equation}
\begin{equation}
B_{ij}({\bf p})=B_0({\bf p})P^{\bot}_{ij}({\bf p})+B_1({\bf
p})P^{||}_{ij}({\bf p}),
\end{equation}
where the projection operators on the transverse and longitudinal
momentum components are given by
\begin{equation}
P^{\bot}_{ij}({\bf p})=\delta_{ij}-\frac{p_ip_j}{{\bf
p}^2},~~~~P^{||}_{ij}({\bf p})=\frac{p_ip_j}{{\bf p}^2}.
\end{equation}
Then the scalar coefficients can be represented as
\begin{equation}
\gamma({\bf p})=\int d{\bf k}\,\omega({\bf p,k})\frac{k_ip_i}{{\bf
p}^2},
\end{equation}
\begin{equation}
\label{B0} B_0({\bf p})=\frac{1}{4}\int d{\bf k}\,\omega({\bf
p,k})\biggl[{\bf k}^2-\frac{(k_ip_i)^2}{{\bf p}^2}\biggr],
\end{equation}
\begin{equation}
\label{B1} B_1({\bf p})=\frac{1}{2}\int d{\bf k}\,\omega({\bf
p,k})\frac{(k_ip_i)^2}{{\bf p}^2},
\end{equation}
where the dynamics enters through the transition rates.

\begin{figure*}
\centering \subfigure{
\resizebox{0.48\textwidth}{!}{%
 \includegraphics{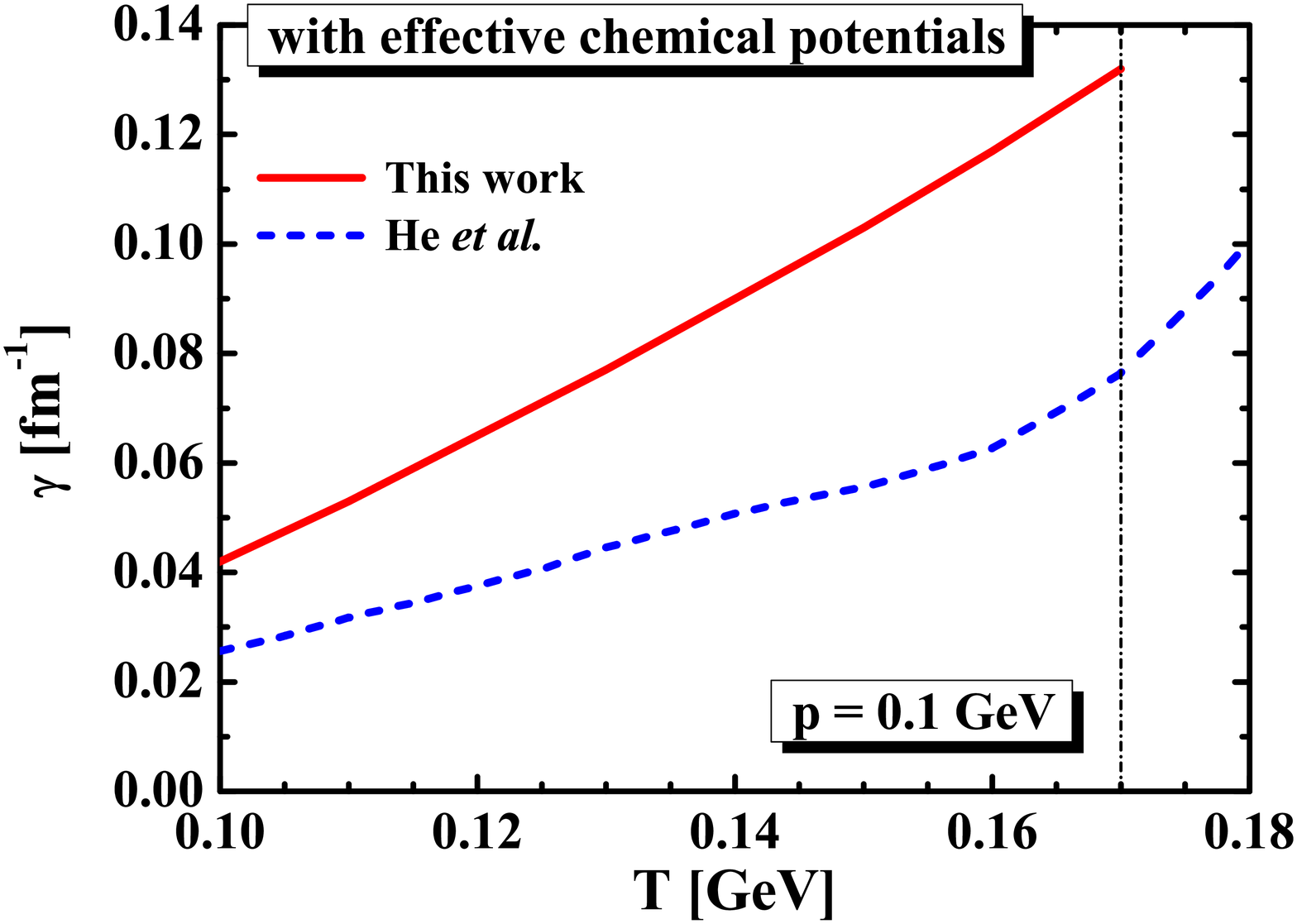}
} } \subfigure{
\resizebox{0.48\textwidth}{!}{%
 \includegraphics{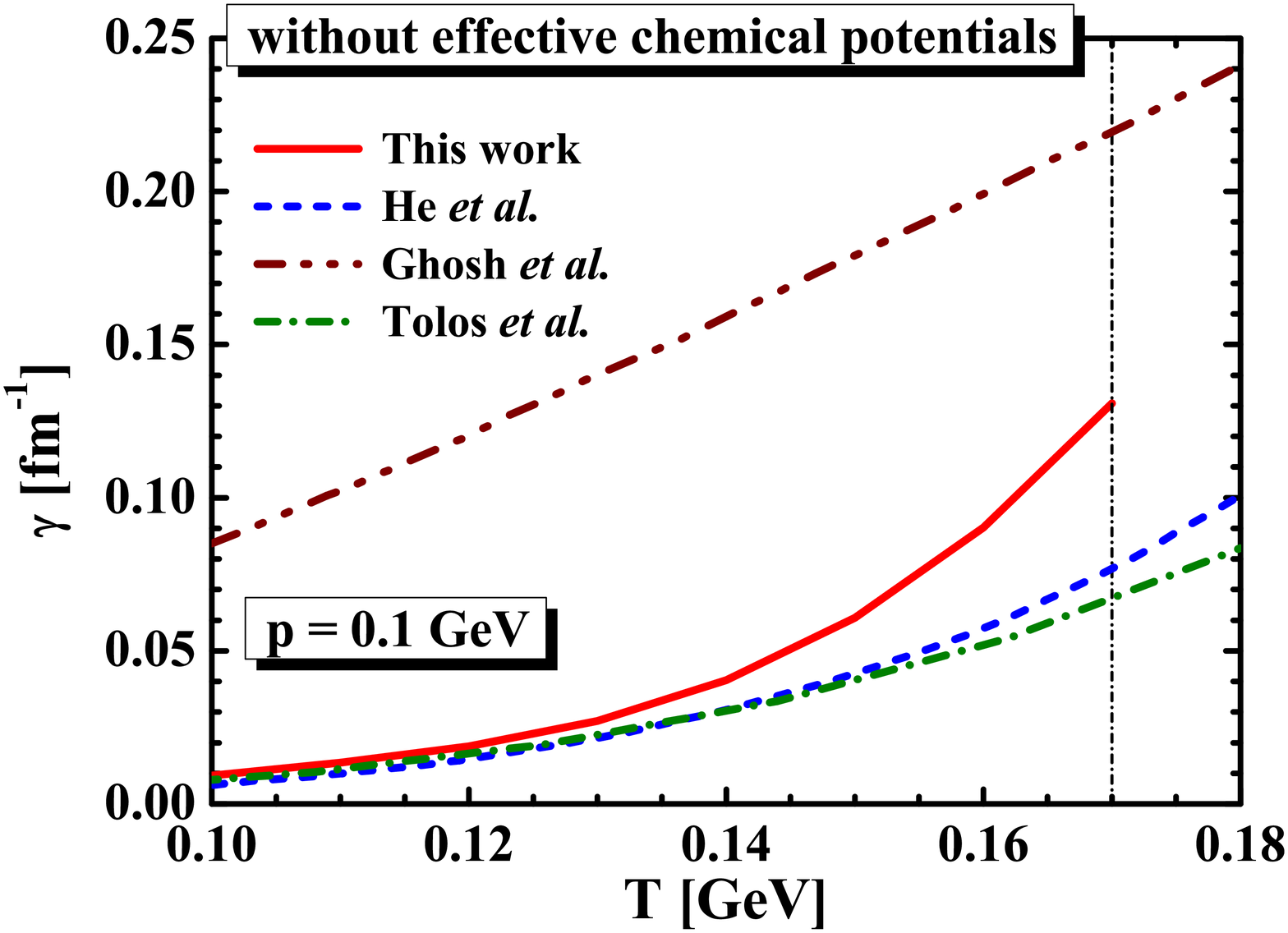}
} } \caption{The thermal relaxation rate of $D$-meson at momentum
$p=0.1~{\rm GeV}$ as a function of temperature with (left panel) and
without (right panel) introducing effective chemical potentials to
the hadronic cocktail. The solid lines represent our results, the
model estimates from Ref.~\cite{diff_hadron_scat_amp} are shown by
the dashed lines, the dash-dot-dotted line depicts the result from
Ref.~\cite{diff_hadron_Born} and the calculation from
Ref.~\cite{diff_hadron_unitarized_3} is presented by the dash-dotted
line.} \label{thermal_relaxation_rate}
\end{figure*}

In order to calculate the $D$-meson drag and diffusion coefficients
in a hot hadron gas, which is discussed in Sec.~\ref{hadron_gas}, we
initialize the $D$-meson in the medium with different initial
momenta ${\bf p}$ at different temperatures. The initial transverse
momentum of the $D$-meson is zero. We implement into the simulation
the cross sections for the  elastic scattering of a $D$-meson with
hadrons as follows
\begin{enumerate}
\item For $D\pi\rightarrow D\pi$, $D\eta\rightarrow D\eta$, and $DK(\bar K)\rightarrow DK(\bar K)$
elastic scattering processes the cross sections are taken from
Ref.~\cite{diff_hadron_unitarized_3}; for $D\rho\rightarrow D\rho$
the cross section is taken from Ref.~\cite{cross_section_Ko}.
\item For $Dm\rightarrow Dm$
elastic scattering processes, where $m$ stands for a meson not
listed above the cross sections are taken as constant,
$\sigma=10~{\rm mb}$, following the calculation of M.~He {\it et
al.}~\cite{diff_hadron_scat_amp}.
\item For $DN(\bar N)\rightarrow DN(\bar N)$ and $D\Delta(\bar\Delta)\rightarrow
D\Delta(\bar\Delta)$ elastic scattering processes, where $N=n,p$ and
$\Delta=\Delta(1232)$, the cross sections are taken from
Ref.~\cite{diff_hadron_unitarized_3}.
\item For $DB\rightarrow DB$ and $D{\bar B}\rightarrow D{\bar B}$
elastic scattering processes, where $B(\bar B)$ is baryon
(antibaryon) not listed above the cross sections are taken as
constant, $\sigma=15~{\rm mb}$, following the calculation of M.~He
{\it et al.}~\cite{diff_hadron_scat_amp}.
\end{enumerate}

For different initial momenta of the $D$-meson we calculate the
following average quantities:
\begin{equation}
\Bigl\langle\frac{dp_z}{dt}\Bigr\rangle,~~\Bigl\langle\frac{dp^2_T}{dt}\Bigr\rangle,~~\Bigl\langle\frac{dp^2_z}{dt}\Bigr\rangle-2p_z\Bigl\langle\frac{dp_z}{dt}\Bigr\rangle,
\end{equation}
which are related to the transport coefficients by
\begin{eqnarray}
\label{transport_coefficient}
A&=&A_z=-\Bigl\langle\frac{dp_z}{dt}\Bigr\rangle,\nonumber\\
B_0&=&\frac{1}{4}\Bigl\langle\frac{dp^2_T}{dt}\Bigr\rangle,\nonumber\\
B_1&=&\frac{1}{2}\biggl[\Bigl\langle\frac{dp^2_z}{dt}\Bigr\rangle-2p_z\Bigl\langle\frac{dp_z}{dt}\Bigr\rangle\biggr].
\end{eqnarray}
We note that   the drag coefficient $A$ in
Eq.~\eqref{transport_coefficient} corresponds to  $A_i({\bf p})$ for
$i=z$, which is given by Eq.~\eqref{drag}.

In Fig.~\ref{drag_and_diffusion} we present the drag and diffusion
coefficients as a function of the momentum of the $D$-meson propagating
through a hadronic medium at different temperatures. We find that
the transport coefficients increase with the momentum of the $D$-meson
and with the temperature of the medium. The dominant contributions
to the drag and diffusion coefficients come from pions, but at
higher temperatures the contributions from other (heavier) hadrons
become important as well.

In addition, we evaluate the thermal relaxation rate of $D$-meson,
which is defined by
\begin{equation}
\gamma=\lim_{p\rightarrow 0}\frac{A}{p}.
\end{equation}
The results of our calculations as well as the comparison with other
previous studies are shown in Fig.~\ref{thermal_relaxation_rate}.
The left panel of Fig.~\ref{thermal_relaxation_rate} presents the
results obtained from our calculation (solid line) in comparison to
the results of Ref.~\cite{diff_hadron_scat_amp} (dashed line) in
which as well effective chemical potentials have been introduced.
The difference between those calculations at lower temperatures is
due to the different cross sections for the $D$-meson scattering
with the hadrons in the medium as well as due to slightly different
effective chemical potentials in both models. At higher temperatures
the difference is a consequence of the larger number of hadron
species in our calculations, necessary to have a critical energy
density of $\varepsilon_c=0.45~{\rm GeV/fm^3}$ that increase the
thermal relaxation rate of $D$-meson. The same differences one finds
in the right panel of Fig.~\ref{thermal_relaxation_rate}, where the
calculations without effective chemical potentials are shown. In
this calculation the pion/proton ratio increases with decreasing
temperature so that at the end of the expansion the hadron gas
consists of $\pi$'s only. The unphysical large value for the
relaxation time obtained in Ref.~\cite{diff_hadron_Born}
(dash-dot-dotted line) comes from the scattering amplitudes for the
$D$-meson interactions with the thermal bath that rapidly grows with
energy and breaks the unitarity condition for the $S$-matrix. In
this work we use the same $D$-hadron cross sections as in
Ref.~\cite{diff_hadron_unitarized_3} and the difference between the
results at higher temperatures is again due to the presence of the
more states in our hadronic cocktail. When we employ the same
hadronic cocktail (pions, kaons, etas, nucleons and
$\Delta$-baryons) as in Ref.~\cite{diff_hadron_unitarized_3}, we
reproduce their results.

In the static limit, where the $D$-meson momentum is going to zero,
one has the property
\begin{equation}
\lim_{p\rightarrow 0}[B_0(p)-B_1(p)]=0,
\end{equation}
which can be obtained from Eqs.~(\ref{B0},\ref{B1}). In this limit
the Einstein relation between the thermal relaxation rate, $\gamma$,
and diffusion coefficient, $B=B_0=B_1$, is fulfilled and defined as
\begin{equation}
\label{Einstein} B=\gamma m_DT,
\end{equation}
where $m_D=1.87~{\rm GeV}$ is the mass of $D$-meson. In this work
the Einstein relation is fully satisfied at lower temperatures, but
starts to deviate slightly with increasing temperature. At the
chemical freeze-out temperature, $T_{\rm ch}=170~{\rm MeV}$, the
deviation is of order of 25\%.

The relaxation time, $\tau_R$, which is defined by
\begin{equation}
\tau_R=\frac{1}{\gamma},
\end{equation}
for $D$-mesons in hadronic matter at the critical temperature,
$T_c$, is approx. equal to $10~{\rm fm}/c$. This value is in a good
agreement with the corresponding value for $c$-quarks in QGP. Thus
the "crossover" constrain is satisfied in our approach.

\begin{figure}
\centering
\includegraphics[width=0.5\textwidth]{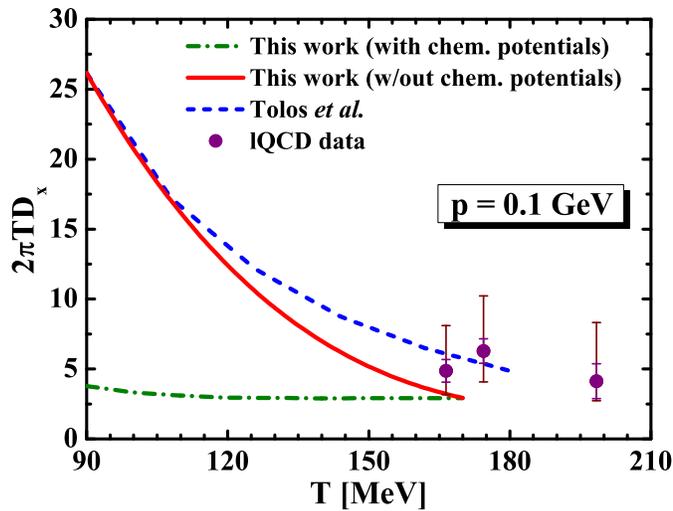}
\caption{The spatial diffusion coefficient of $D$-meson at momentum
$p=0.1~{\rm GeV}$ as a function of temperature with (dash-dotted
line) and without (solid line) the introduction of effective
chemical potentials to the hadronic cocktail in comparison to the
results from Ref.~\cite{diff_hadron_unitarized_3} (dashed line). The
lQCD data are taken from Ref.~\cite{lQCD_results}
(symbols).}\label{spatial_diffusion}
\end{figure}

Finally, we calculate the spatial diffusion coefficient, $D_x$,
which describes the broadening of the spatial distribution with
time,
\begin{equation}
\langle{\bf x}^2(t)\rangle-\langle{\bf x}(t)\rangle^2\simeq 6D_xt,
\end{equation}
and it can be related to the thermal relaxation rate and diffusion
coefficient through
\begin{equation}
D_x=\lim_{p\rightarrow 0}\frac{B}{m_D^2\gamma^2}.
\end{equation}
In Fig.~\ref{spatial_diffusion} we present the results for the
spatial diffusion coefficient calculated in our approach with
(dash-dotted line) and without (solid line) the introduction of the
effective chemical potentials to the hadronic cocktail in comparison
to the results from Ref.~\cite{diff_hadron_unitarized_3} (dashed
line), where the effective chemical potentials are not included. At
the critical temperature there is a very smooth transition between
our calculations for the hadron gas and the lQCD results taken from
Ref.~\cite{lQCD_results}. The difference between the results at
higher temperatures shown by solid and dashed lines can be explained
by the presence of the higher states in our hadronic cocktail.

\section{The nuclear modification factor and elliptic flow of $D$-mesons}
\label{R_AA_and_v2}

The size of the transport coefficients in hadronic matter shows that
the hadronic contributions should be included when evaluating the
nuclear suppression factor and the elliptic flow of $D$-mesons as
well as of single nonphotonic electrons.

We implement the transport coefficients calculated in the hadronic
matter with the introduction of effective chemical potentials
(Fig.~\ref{drag_and_diffusion}) to the MC@sHQ model and calculate
the nuclear modification factor, $R_{AA}$, and elliptic flow, $v_2$,
for two different scenarios:
\begin{enumerate}
\item {\it scenario I:} we use the transport
coefficients, drag and diffusion coefficients, obtained from
Eq.~\eqref{transport_coefficient}, however, in this case it is not
evident that the asymptotic solution of the FP
equation~\eqref{FP_equation} is the thermal equilibrium distribution
function.
\item {\it scenario II:} we employ the drag coefficient, obtained
from Eq.~\eqref{transport_coefficient}, and assume, $B=B_0=B_1$,
calculated by Eq.~\eqref{Einstein}. Thus we impose the Einstein
relation which is valid only for small momenta. The Einstein
relation ensures us that the asymptotic distribution function is the
Boltzmann distribution function.
\end{enumerate}
We consider both scenarios I and II to provide an idea of the
theoretical systematic uncertainties of the FP approximation to the
full Boltzmann equation.

\begin{figure*}
\centering
\includegraphics[width=\textwidth]{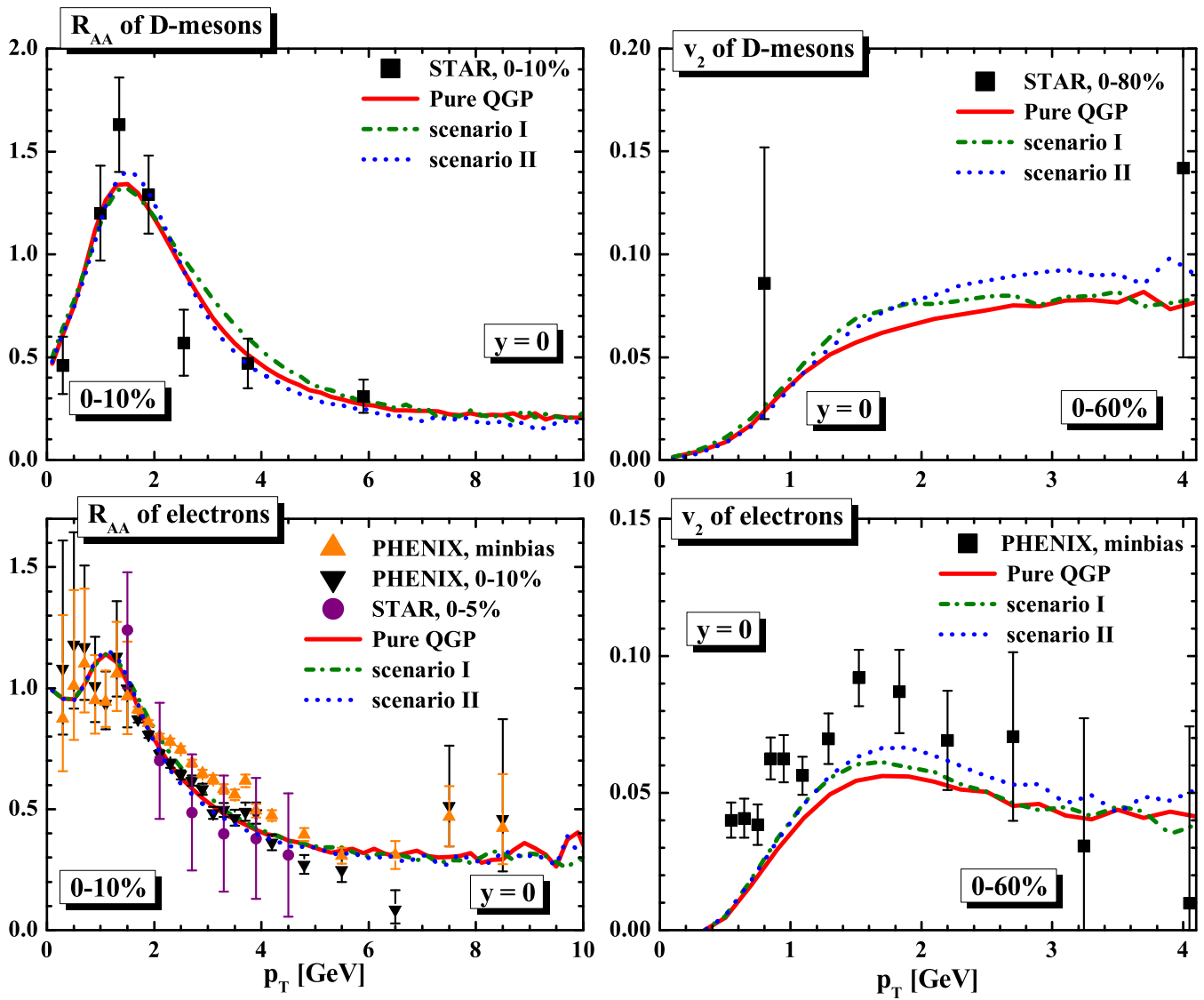}
\caption{The nuclear modification factor, $R_{AA}$, and elliptic
flow, $v_2$, of $D$-mesons and of single nonphotonic electrons
originating from the decays of HF mesons for two scenarios as well
as for pure QGP scenario in comparison to the experimental data from
STAR and PHENIX experiments: $R_{AA}$ of $D$-mesons -
Ref.~\cite{STAR_Dmeson_R_AA}, $v_2$ of $D$-mesons -
Ref.~\cite{STAR_Dmeson_v2}, $R_{AA}$ and $v_2$ of decay electrons -
Ref.~\cite{PHENIX_electrons_R_AA_v2,STAR_electrons_R_AA}.}
\label{R_AA_v2}
\end{figure*}

In Fig.~\ref{R_AA_v2} we show the results for $R_{AA}$ and $v_2$ of
$D$-mesons and of single nonphotonic electrons as a function of the
transverse momentum. The comparison with the experimental data is
presented, too. According to these results we can conclude that the
$D$-meson rescattering in the hadronic gas is almost invisible for
the $R_{AA}$ because of the compensation between the extra cooling
and extra radial flow, but shows a systematic increase of 1\%-2\% to
the $v_2$ of $D$-mesons and of decay electrons. The results of other
groups~\cite{HF_Linnyk,He_RAA_v2} also indicate the important role
of the hadronic matter in the suppression and elliptic flow of HF
mesons.

\section{The $D$-meson transport coefficients at FAIR energies}
\label{FAIR}

\begin{figure*}
\centering \subfigure{
\resizebox{0.48\textwidth}{!}{%
 \includegraphics{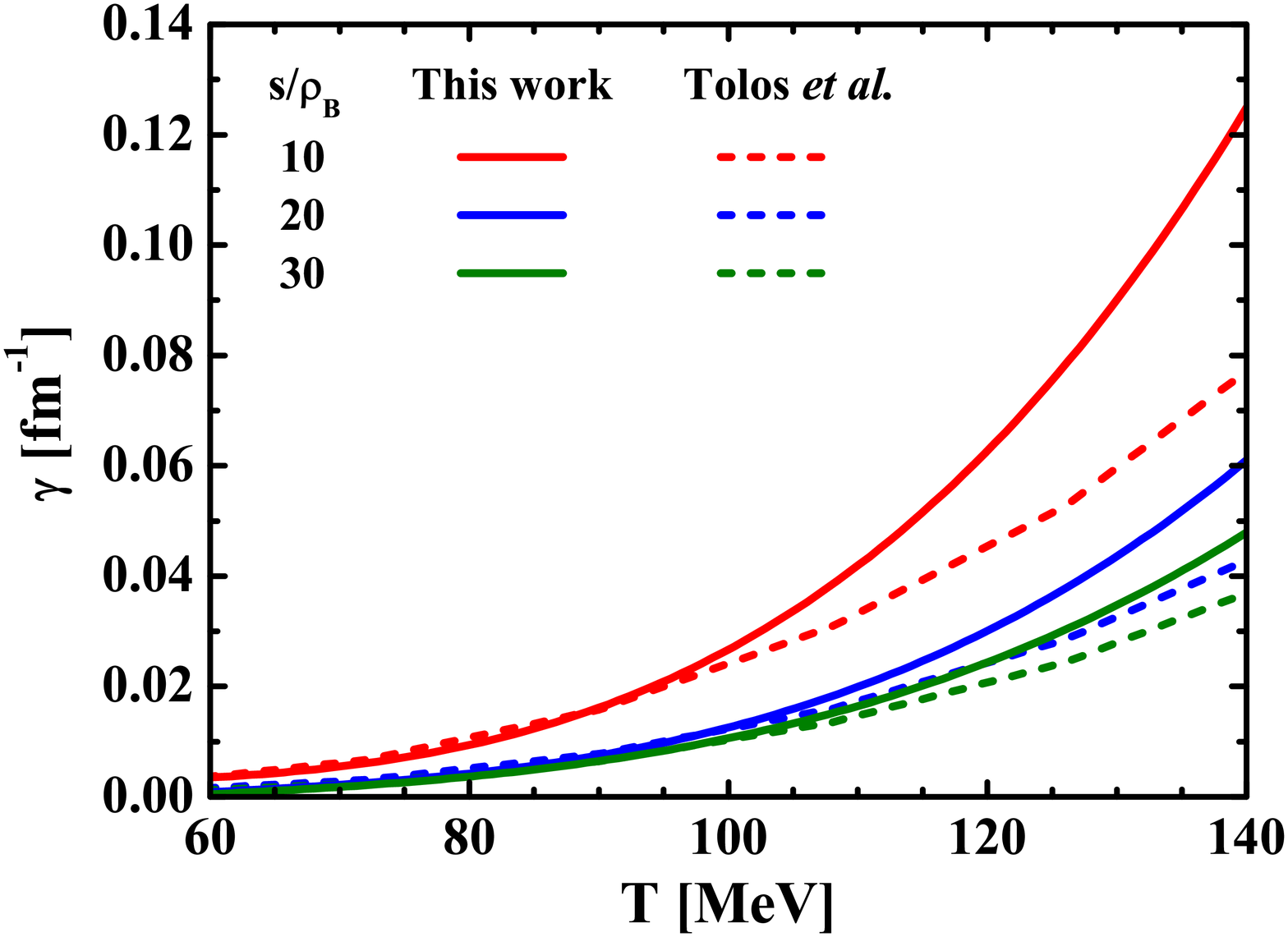}
} } \subfigure{
\resizebox{0.48\textwidth}{!}{%
 \includegraphics{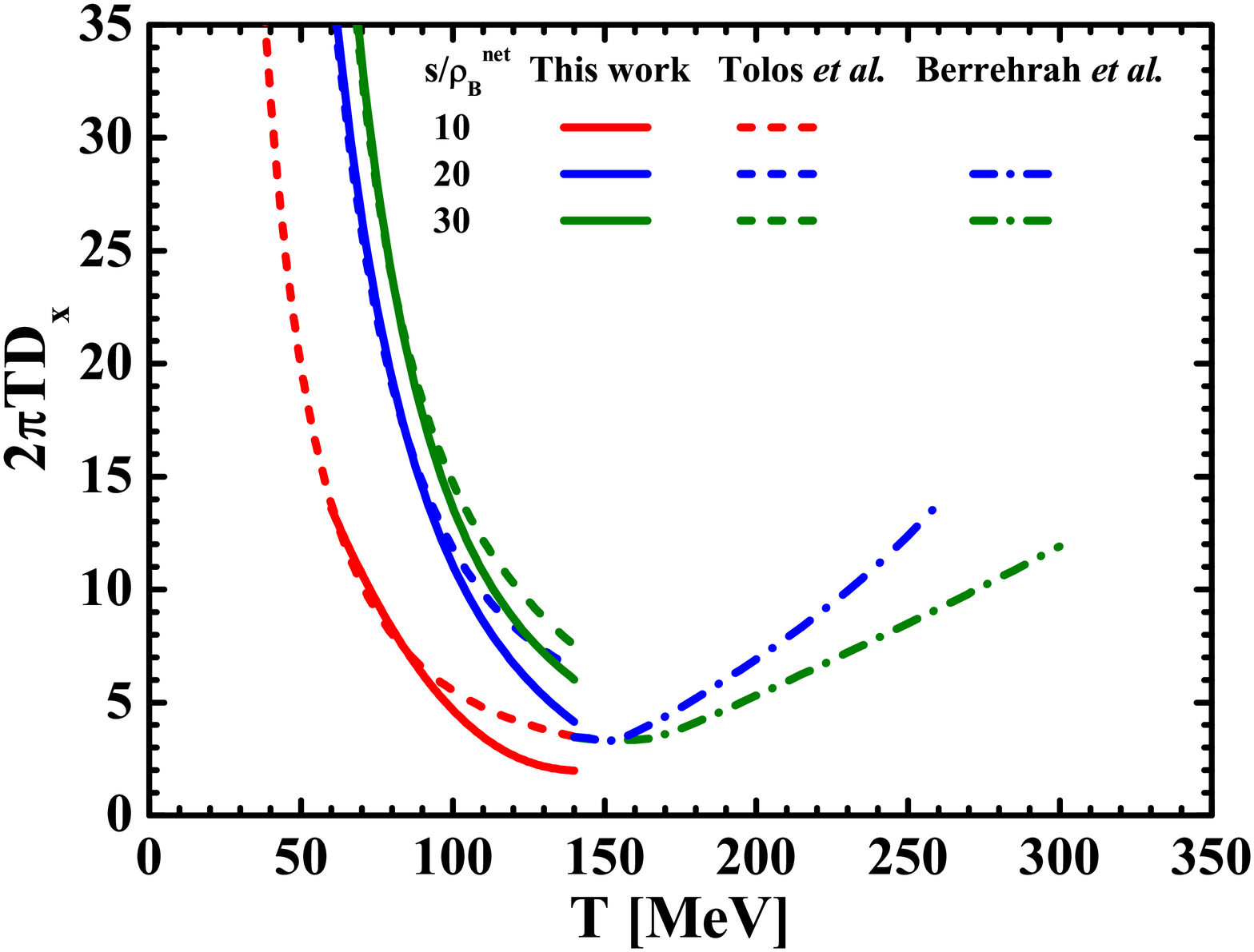}
} } \caption{The thermal relaxation rate (left panel) and the
spatial diffusion coefficient (right panel) of $D$-meson for three
different isentropic trajectories as a function of temperature in
comparison to the model estimates from
Ref.~\cite{diff_hadron_unitarized_3} (dashed lines). The results for
the spatial diffusion coefficient of $c$-quarks propagating in the
partonic medium obtained in Ref.~\cite{spatial_Hamza} are shown by
the dash-dotted lines.} \label{transport_FAIR}
\end{figure*}

In this section we construct the hadronic isentropic trajectories,
i.e., with fixed specific entropy density, appropriate at future
FAIR and NICA heavy-ion experiments. In particular, for FAIR
physics, the beam energy will run from $\sqrt{s}=5-40~{\rm AGeV}$,
which approximately corresponds to $s/\rho_B^{\rm
net}=10-30$~\cite{Bravina}. We take three characteristic values of
$s/\rho_B^{\rm net}=10,20,30$ and show in Fig.~\ref{transport_FAIR}
the results for the thermal relaxation rate (left panel) and the
spatial diffusion coefficient (right panel) of $D$-meson for these
trajectories as a function of temperature in comparison to the model
estimates from Ref.~\cite{diff_hadron_unitarized_3}.
%We note that
%the spatial diffusion coefficient, $D_x$, which describes the
%broadening of the spatial distribution with time,
%%
%\begin{equation}
%\langle{\bf x}^2(t)\rangle-\langle{\bf x}(t)\rangle^2\simeq 6D_xt,
%\end{equation}
%%
%can be related to the thermal relaxation rate and diffusion
%coefficient through
%%
%\begin{equation}
%D_x=\lim_{p\rightarrow 0}\frac{B}{m_D^2\gamma}.
%\end{equation}
%
We find that the $D$-meson thermal relaxation rate, $\gamma$, and
the spatial diffusion coefficient, $D_x$, strongly depend on the
isentropic trajectory. The impact of baryons to the transport
coefficients becomes more important due to the large values for the
baryon chemical potential corresponding to the isentropic
trajectories, which are used. We remind that we use the same cross
sections for $D$-meson rescattering with hadrons in the medium as in
Ref.~\cite{diff_hadron_unitarized_3}. The difference between
presented results is due to the presence of the higher states in our
hadronic cocktail as they play a significant role at higher
temperatures. Up to know, the approach in
Ref.~\cite{FP_Gossiaux_Aichelin} was not extended to the finite
chemical potentials. We therefore compare our results with the
results for the spatial diffusion coefficient of $c$-quarks
propagating in the partonic medium calculated in
Ref.~\cite{spatial_Hamza}. The comparison is presented in the right
panel of Fig.~\ref{transport_FAIR}. For the isentropic trajectory
with $s/\rho_B^{\rm net}=20$ there is a perfect matching of results
for $D$-mesons and $c$-quarks propagating through the hadronic and
partonic matter, respectively.

\section{Summary}
\label{summary} Heavy mesons, created in ultrarelativistic heavy-ion
collisions, are presently considered as a very good observable to
improve our knowledge of the time evolution of the QGP, formed in
these collisions. Being produced in hard collisions with no initial
elliptic flow their thermalization time is considerably larger than
the expansion time of the QGP. Therefore the final spectra of heavy
mesons as well as their elliptic flow carry information on the
interaction of the heavy quarks with the QGP partons. In order to
obtain a quantitative evaluation of the modification of the heavy
quark momentum distribution due to collisions with QGP particles the
influence of hadronic rescattering of heavy mesons on the spectra
has to be studied. Employing drag and diffusion coefficients based
on the most recent calculations of the cross section of $D$-mesons
with the hadronic environment, we use the FP equation to determine
the importance of the hadronic rescattering on the heavy meson
spectra. We find that the influence on $R_{AA}$ is negligible
because of the compensation of the extra cooling and extra radial
flow, whereas for the elliptic flow the influence is visible but
small as compared to the elliptic flow the heavy quarks acquire
during the expansion of the QGP.

We also extend our calculations to the finite chemical potentials,
appropriate at future FAIR and NICA experiments, and show the
$D$-meson spatial diffusion coefficient. We find that for the
isentropic trajectory with $s/\rho_B^{\rm net}=20$ there is a
perfect matching of our results with the results for $c$-quarks
propagating in the partonic medium calculated in
Ref.~\cite{spatial_Hamza}.

%The calculation of the drift coefficient allows to calculate the spatial diffusion coefficient. Our calculation shows that close to $ T_c$
%the spatial diffusion coefficient is close to the value obtained with help of  lattice gauge calculations. This confirms that
%the cross sections employed to calculate the drift coefficient.

\section*{Acknowledgments}
We thank E.~L.~Bratkovskaya, H.~Berrehrah, D.~Cabrera and H.~ van
Hees for fruitful discussions. We also thank E.~L.~Bratkovskaya for
providing us the code to simulate the hadron matter. This work was
supported by the Project TOGETHER (Pays de la Loire) and
I3-Hadronphysics II. V.O. acknowledges financial support from the
HGS-HIRe Program. L.T. acknowledges support from the Ramon y Cajal
Research Programme from Ministerio de Ciencia e Innovaci\'on and
from Ministerio de Ciencia e Innovaci\'on under contracts
FPA2010-16963 and FPA2013-43425-P.

%\begin{eqnarray}
%&&\hspace{-1.0cm}\omega({\bf p,k})=\gamma_h\int\frac{d{\bf
%q}}{(2\pi)^9}\frac{f_h({\bf q})[1\pm f_h({\bf
%q+k})]}{(2E^D_p)(2E^h_q)(2E^D_{p-k})(2E^h_{q+k})} \nonumber\\
%&&\hspace{0.5cm}\times\,(2\pi)^4\delta(E^D_p+E^h_q-E^D_{p-k}-E^h_{q+k})|M|^2.
%\end{eqnarray}
%
%Here $h$ stands for the hadrons in the thermal bath, $\gamma_h$
%represents the spin-isospin degeneracy factor and $f_h$ are the Bose
%or Fermi distributions for thermal hadrons. The Bose enhancement
%factor for mesons, $1+f_h({\bf q+k})$, and the Pauli blocking term
%for baryons, $1-f_h({\bf q+k})$, are included.

\end{document}